\documentstyle[12pt]{article}

\def\bib{\bibitem}
\def\be{\begin{equation}}
\def\ee{\end{equation}}
\def\barr{\begin{array}}
\def\earr{\end{array}}
\def\dis{\displaystyle}

\def\etal{ {\em et al.}}
\def\ie{ {\em i.e.}}
\def\viz{ {\em viz.}}

\def\gev{\: {\rm GeV} }

\def\pb{\: {\rm pb}}
\def\fb{\: {\rm fb}}
\def\ra{\rightarrow}
\def\mand{\qquad {\rm and} \qquad}
\def\ptsl{p_T \hspace{-1.1em}/\;}
\def\pslash{p \hspace{-0.6em}/\;}
\def\msl{m \hspace{-0.8em}/\;}

\def\kg{\kappa_\gamma}

\def\dkg{\Delta \kappa_\gamma }
\def\dkZ{\Delta \kappa_Z }
\def\dkgz{\Delta \kappa_{\gamma,Z} }
\def\lg{\lambda_\gamma}
\def\lZ{\lambda_Z}
\def\lgZ{\lambda_{\gamma,Z}}
\def\dgZ{\Delta g_1^Z}

\def\bdkZ{\overline{\Delta \kappa_Z} }
\def\blZ{\overline{\lambda_Z}}
\def\bdgZ{\overline{\Delta g_1^Z}}

\def\ib#1,#2,#3{       {\it ibid.\/ }{\bf #1} (19#2) #3}
\def\ap#1,#2,#3{       {\it Ann.~Phys.~(NY)\/ }{\bf #1} (19#2) #3}
\def\ijmp#1,#2,#3{     {\it Int.~J.~Mod.~Phys.\/ } {\bf A#1} (19#2) #3}
\def\mpl#1,#2,#3 {     {\it Mod.~Phys.~Lett.\/ } {\bf A#1} (19#2) #3}
\def\np#1,#2,#3{       {\it Nucl.~Phys.\/ }{\bf B#1} (19#2) #3}
\def\npps#1,#2,#3{     {\it Nucl.~Phys.~B (Proc.~Suppl.)\/ }{\bf B#1}
                             (19#2) #3}
\def\plb#1,#2,#3{      {\it Phys.~Lett.\/ }{\bf B#1} (19#2) #3}
\def\pr#1,#2,#3{       {\it Phys.~Rev.\/ }{\bf #1} (19#2) #3}
\def\prd#1,#2,#3{      {\it Phys.~Rev.\/ }{\bf D#1} (19#2) #3}
\def\prep#1,#2,#3{     {\it Phys.~Rep.\/ }{\bf #1} (19#2) #3}
\def\prl#1,#2,#3{      {\it Phys.~Rev.~Lett.\/ }{\bf #1} (19#2) #3}
\def\pro#1,#2,#3{      {\it Prog.~Theor.~Phys.\/ }{\bf #1} (19#2) #3}
\def\rmp#1,#2,#3{      {\it Rev.~Mod.~Phys.\/ }{\bf #1} (19#2) #3}
\def\sp#1,#2,#3{       {\it Sov.~Phys.-Usp.\/ }{\bf #1} (19#2) #3}
\def\zpc#1,#2,#3{      {\it Zeit.~f\"ur Physik\/ }{\bf C#1} (19#2) #3}
\def\appb#1,#2,#3{     {\it Acta Phys.\ Polon.\/ }{\bf B#1} (19#2) #3}

\begin{document}
\thispagestyle{empty}
\setcounter{page}{0}
\renewcommand{\thefootnote}{\fnsymbol{footnote}}

\begin{flushright}
MPI-PTh/96-73\\[1.7ex]
IFT-96/17\\[1.7ex]
{\large \tt hep-ph/9608416} \\
\end{flushright}

\vskip 45pt
\begin{center}
{\Large \bf Unravelling the {\boldmath $WW\gamma$ } and
            {\boldmath $WWZ$} Vertices at the \\[1.5ex]
            Linear Collider: {\boldmath $\bar{\nu} \nu\gamma$ }
            and {\boldmath $\bar{\nu}\nu\bar{q}q$}
final states}

\vspace{11mm}
{\large Debajyoti Choudhury}\footnote{debchou@mppmu.mpg.de}\\[1.1ex]
{\em Max--Planck--Institut f\"ur Physik,
              Werner--Heisenberg--Institut,\\
              F\"ohringer Ring 6, 80805 M\"unchen,  Germany.}\\[2ex]
and\\[2ex]
{\large Jan Kalinowski}\footnote{kalino@fuw.edu.pl}
                 \footnote{Supported in part by the Polish Committee for
                            Scientific Research.}\\[1.1ex]
{\em Institute of Theoretical Physics, Warsaw University,\\
Ho\.za 69, 00-681 Warsaw, Poland.} \\[2ex]

\vspace{50pt}
{\bf ABSTRACT}
\end{center}
\begin{quotation}
   We perform a detailed analysis of the processes $e^+e^- \rightarrow
\bar{\nu} \nu\gamma$ and  $\bar{\nu}\nu\bar{q}q$ at future linear $e^+e^-$
colliders and assess
their sensitivity to anomalous gauge boson couplings. We
consider center of mass energies $\sqrt{s}=$ 350, 500 and 800 GeV.
We demonstrate that significant improvements can be obtained if the
phase space information for the cross sections is used maximally.
At 800 GeV the parameters $\Delta\kappa_{\gamma}$ and $\lambda_{\gamma}$
can be constrained, at 95\% CL, to about 0.02 and 0.01, while
the parameters $\Delta\kappa_Z$, $\lambda_Z$ and $\Delta g_1^Z$ can be
probed down to about 0.009, 0.002 and 0.004  respectively.
The precision of these measurements is likely to be limited
by statistical errors at anticipated luminosities at these energies.
\end{quotation}

\newpage
\renewcommand{\thefootnote}{\arabic{footnote}}

\section{Introduction}
     \label{sec:introd}
Although the LEP and SLC measurements are often claimed to have
vindicated the Standard Model (SM) at better than 1\% level~\cite{lep-slc},
in reality, this success has been limited to an accurate determination
of the fermion-vector boson couplings. The nature of the vector boson
self couplings, as yet, is only poorly measured. Thus, inspite of
our prejudice for the `gauge dogma', we are still far from establishing
the $W$ and the $Z$ to be gauge bosons. On the other hand, even if we
are to believe in the SM, we are led to another question. Is there a
``great desert'' beyond, or does new physics soon overwhelm the SM?
Can (all) the particles associated with the new physics
be discovered directly ? If not, can the existence of new thresholds
cause significant departures in the vector boson self couplings?
This seems quite likely in the light of the remarkable agreement
between the direct measurement of the top quark mass and the
indications from the
precision measurements at LEP. A precise measurement
of these vertices is thus of paramount importance.

Departures from the gauge theory prediction are, perhaps, best constrained
from a consideration of the virtual effects that these engender
in various low energy observables like the anomalous magnetic
moment of the muon~\cite{mu_anom}, rare meson decays~\cite{meson_decay}
and the precision electroweak data~\cite{oblique}. However, such bounds
suffer from the fact that they necessitate either an assumption of
lack of any {\em cancellation} between various corrections, or
a complete specification of the new physics scenario.

It is thus more attractive to measure such effects directly and, indeed,
much effort has been directed towards this goal, both in the context
of the forthcoming LEP2 operation~\cite{LEP2}, or the proposed
Linear Colliders (LC)~\cite{LC_review}.
The $W$ pair-production process has been
studied in great detail as a probe for gauge boson self-couplings.
In this article we argue
that an additional set of measurements, at no extra cost, can and
{\em should} be made at the LC in our quest of unravelling this
sector of the SM.
One of these ($e^+e^-\rightarrow \bar{\nu} \nu\gamma$)  is
sensitive solely to the $W^+W^-\gamma$ vertex while the other
($e^+e^- \ra \bar\nu \nu \bar{q} q $) is sensitive primarily
to the $W^+W^-Z$ coupling.
With the aditional knowledge that these measurements
will provide, a {\em model independent} evaluation would be much easier
to achieve.

In the next section, we review the generic triple vector boson vertex
(TGV) and summarize current experimental limits and improvements that
can be achieved at LEP2 and Tevatron. In Sec. 3 we argue for the case
of $\bar{\nu}\nu\gamma$ and $\bar{\nu}\nu Z$  final states.
They are then discussed in detail and expected experimental bounds
are derived  in the following two sections. Finally we
summarize our conclusions in Sec. 6.

\section{The Anomalous Couplings}
         \label{sec:anomcoupl}
New physics can, and, in general, will manifest itself in
corrections to both the triple and the quartic vector boson vertex.
We concentrate here, however, only on the former set.
While quartic couplings can be investigated at the
LC~\cite{LC_review,quartic},
the expected bounds tend to be weaker. Furthermore, in an
effective Lagrangian language, any operator
that would cause a modification in the quartic vertex without causing
a corresponding one in the TGV, would be
suppressed by higher powers of momentum.

Expressed in  purely phenomenological terms, the effective Lagrangian
for the $WW\gamma$ and the $WWZ$ vertex can be expressed
in terms of seven parameters each~\cite{lagr_effec}. Of these seven,
three violate $CP$ while a fourth one violates $C$ and $P$ but
conserves $CP$. As, pending large cancellations, data on the neutron
electric dipole moment constrain these very severely~\cite{edmn}, we
shall not discuss them any
further\footnote{In a collider, these operators are best
          isolated by looking at final state
          asymmetries~\protect\cite{biswarup}.}.
The effective Lagrangian, thus restricted, can be expressed
as (with $V \equiv \gamma$ or $Z$)
\be
\barr{rl} \dis
     {\cal L}_{\it eff}^{WWV} = &
      \dis -i g_{V} \Bigg[ \hspace*{0.3em}
                         ( 1 + \Delta g^V_1 )
                           \left( W^\dagger_{\alpha \beta} W^\alpha
                                 - W^{\dagger\alpha} W_{\alpha \beta}
                           \right) V^\beta
                         +
                          ( 1 + \Delta \kappa_V)
                            W^\dagger_{\alpha} W_\beta V^{\alpha\beta}
              \\[1.5ex]
            & \hspace*{2.3em} \dis
               + \frac{\lambda_V}{M_W^2}
                 W^\dagger_{\alpha \beta} {W^\beta}_\sigma
                 V^{\sigma\alpha} \Bigg]
\earr
      \label{lagrangian}
\ee
where $V_{\alpha\beta} = \partial_\alpha V_\beta - \partial_\beta
V_\alpha $ and  $W_{\alpha\beta} = \partial_\alpha W_\beta -
\partial_\beta W_\alpha $.
In eq.~(\ref{lagrangian}), $g_V$ is   the overall $WWV$ coupling in
the SM, \viz,
\be
    g_\gamma = e, \qquad g_Z = e \cot \theta_W \ ,
\ee
where $\theta_W$ is the weak mixing angle. Electromagnetic gauge
invariance requires that $\Delta g^\gamma_1 (q^2 = 0) =0 $, though it can
assume other values for off-shell photons, a fact often missed in
the literature. Within the SM, we have, at the tree level,
\be
\Delta g_1^\gamma = \dgZ = \dkg = \dkZ = \lg = \lZ = 0 \ .
\ee
It is easy to see that eq.(\ref{lagrangian}) is the most general
expression consistent with Lorentz, $C$ and $P$ invariance. All
higher derivative terms can be reabsorbed into the couplings above
provided they are treated as form-factors and not constants. It is thus
important to bear in mind the fact that the strength of the various
terms in the vertex would vary (in general, independently) with the
momentum scale of the process being considered.

To date, the only direct constraints are those obtained for
$\kg$ and $\lg$ from an analysis of the $W \gamma$ events at
the Tevatron~\cite{tevatron}. The 95\% C.L. bounds are
\be
    -1.6 < \dkg < 1.8 \mand -0.6 < \lg < 0.6 \ .
            \label{tev_lim}
\ee
It is also
estimated that an analysis of the $WW$ and $WZ$ events in the
Run 1b data will lead to bounds like $-0.65 < \dkg < 0.75 $ and
$ -0.4 < \lambda_{\gamma,Z} < 0.4$. Further improvements
(approximately by a factor of 2) are expected if the main
injector becomes operative~\cite{main_inject}.

In the context of the oncoming runs at LEP2, considerable
efforts~\cite{LEP2} have been made to determine the
sensitivity of $W^+ W^-$ pair production (rate as well as phase space
distribution) on the anomalous couplings with the
conclusion: the
individual bounds would be quite similar to those obtainable at the
Tevatron, though the parameter space contours would be different.
While LEP2 {\em might} do marginally better for $\dkgz$, Tevatron
is better poised\footnote{Since these represent higher dimensional
                      operators, their contribution to the cross section
                      increases with the center of mass energy, unless
                      the couplings themselves decrease with the
                      momentum transfer.}
for constraining $\lgZ$. At the LC however, the same process would
lead to more than an order of magnitude improvement in the
bounds~\cite{LC_review}.
Only then would we begin to probe the radiative corrections to these
couplings expected within the SM~\cite{SM_rad} or the minimal
supersymmetric standard model (MSSM)~\cite{MSSM_rad}.

\section{The Case for {\boldmath $  \bar{\nu} \nu \gamma$}
         and {\boldmath $  \bar{\nu} \nu Z$} Final   States}
       \label{sec:case}
It is quite obvious that compared to $W$-pair production,
the processes being proposed here are suppressed by higher
powers of $g_V$.
Thus the cross sections are, naively, expected to be smaller.
Furthermore, the invisibility
of the neutrinos lead to considerable loss of information. This, then,
points to a reduced sensitivity. Why therefore
consider such final states at all?
To appreciate this, one needs to recount that a process like
$f \bar{f} \ra W^+ W^-$ receives contribution from both the $\gamma$--
and the $Z$--mediated $s$-channel process. Thus, all of the
six\footnote{Since the photon is off-shell, $g_1^\gamma$ is no longer
          constrained to be unity.}
$WW\gamma/Z$ couplings are inexorably intertwined. This leads to two
possible difficulties. For one, it is quite conceivable that the new
physics corrections to the couplings are such
that their contributions to the $W$ pair production
cancel to a significant degree, thus reducing sensitivity.
On the other hand, even if a deviation in the cross section is seen,
this would not point us to a unique theory. These are critical issues
as a measurement of the TGV at the LC could be a
critical tool in examining radiative corrections within the SM or the
MSSM.

There are two solution to the imbroglio. One is to consider experiments
in different collider modes and combine the results obtained therein.
For example, the cross sections for both $W\nu$ production
in an $e \gamma$ collider~\cite{photon}
and $W$-pair production in $\gamma \gamma$ colliders~\cite{photon}
are independent of the $WWZ$ vertex (as well as the form factor nature of
$g_1^\gamma$). On the other hand, all the couplings do contribute
to $We\nu$ production at an  $e^- e^-$ collider. However, the interplay
being different, this mode can provide significant new information~\cite{ee}.
If one were not willing to consider another machine, the alternative
would be to look for processes with different dependence on the
anomalous couplings. Hadronic colliders have an advantage in that they
permit many more potentially interesting modes. We already have mentioned
three in the context of the Tevatron. Unfortunately, the sensitivity
tends to be lower for the noncanonical final states.

At the canonical LC, inspite of the lower sensitivity expected, it is thus
interesting to consider final states other than $W^+ W^-$. The two
simplest examples of this are provided by $\bar{\nu} \nu \gamma$ (sensitive
only to $\dkg$ and $\lg$, $g_1^\gamma$ being unity as the photon is on-shell)
and $\bar{\nu} \nu Z$ (sensitive to $\dkZ$, $\lZ$ and $\dgZ$). Although both
these processes have already been considered in the
literature~\cite{godfrey,abraham,ambro-mele,LEP2}, we bring into the analysis
a greater degree of sophistication and demonstrate that such an
experiment is actually much more sensitive than what was previously
thought.

\section{\boldmath $e^+ e^- \ra \bar \nu \nu \gamma$}
     \label{sec:nngamma}
The signal here would be a single energetic photon accompanied by
missing momentum (equal and opposite to that of the photon). While
a significant fraction of such events would arise from $Z \gamma$
production, their importance can be reduced by appropriate kinematical
cuts, as we shall demonstrate below.
\begin{figure}[h]
        \vskip 4in\relax\noindent\hskip -1.8in
        \relax{\includegraphics{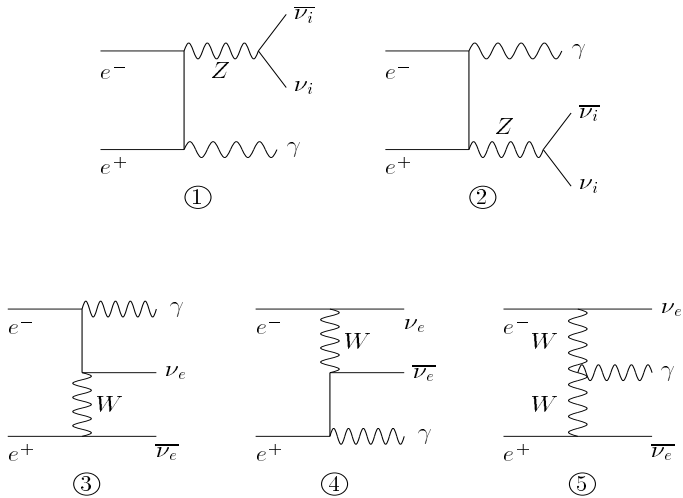}}
        \vspace{-20ex}
 \caption{{\em The Feynman diagrams responsible for
          $e^+ e^- \ra \bar \nu \nu \gamma$.}}
 \label{fig:feyn_nng}
\end{figure}
The relevant Feynman diagrams are shown in
Fig.~\ref{fig:feyn_nng}. Diagrams (1 \& 2) form a gauge-invariant
subset by themselves, and also lead to $\bar \nu_\mu \nu_\mu \gamma$
and $\bar \nu_\tau \nu_\tau \gamma$ final states. The helicity
amplitudes can be found in ref.~\cite{abraham}. Before examining
the cross-sections, let us delineate first the kinematical cuts that
we will employ. It is clear that bremsstrahlung-type diagrams lead,
preferentially, to photons close to the beam axis. To enhance the
importance of the TGV, it is thus necessary to
impose an angular cut on the photon. We find
\be
   25^\circ < \theta_\gamma < 155^\circ
               \label{cut:th_gam}
\ee
to be a suitable choice.
Further, the photon must have sufficient energy to be
detectable. We thus require
\be
   E_\gamma > 25 \gev\ .
               \label{cut:e_gam}
\ee
The above combination also implies that the events must have a
missing transverse momentum $\ptsl > 10.6 \gev$. This then takes care
of the process $e^+ e^- \ra \gamma \gamma$ with one photon disappearing
into the beam pipe.
\begin{figure}[h]
        \vskip 4.5in\relax\noindent\hskip -1.8in
        \relax{\includegraphics{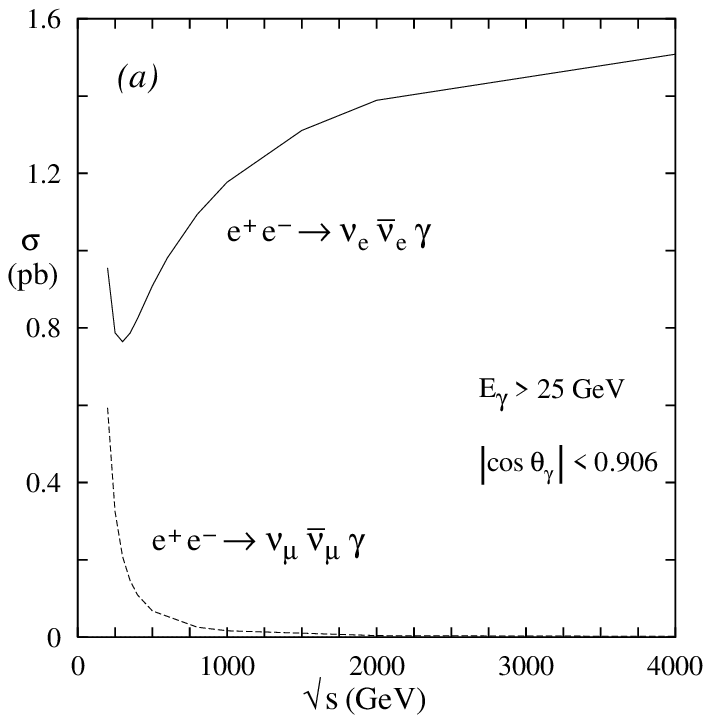}}
       \relax\noindent\hskip 3in
              \relax{\includegraphics{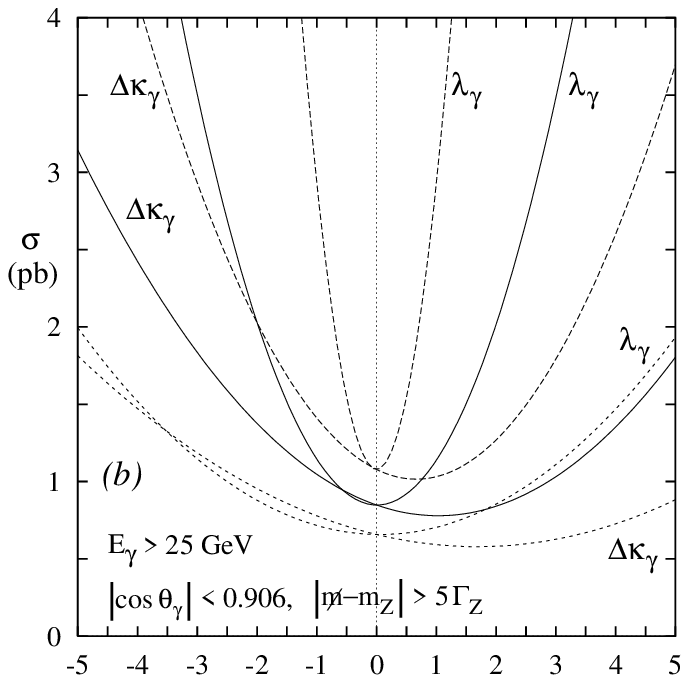}}
        \vspace{-20ex}
 \caption{{\em {\em (a)} The variation of the SM cross sections with the
               center of mass energy; and
               {\em (b)} The total single photon
               cross section as a function of the individual
               anomalous couplings for a fixed center of mass
               energy (dotted : 350 GeV, solid : 500 GeV, dashed : 800 GeV).
               The other coupling is held to be vanishing.
              }}
 \label{fig:cs_gam}
\end{figure}
The SM production cross section with these cuts is presented in
Fig.~\ref{fig:cs_gam}{\em a}. The increase in the rates with
$\sqrt{s}$ is symptomatic of the diagram with the
TGV\footnote{Of course, gauge invariance does ensure partial
         wave unitarity, as can be explicitly checked.}. Also
shown in the figure is the cross-section for
$\bar \nu_\mu \nu_\mu \gamma$ production. At a first glance,
it might seem that the latter is an unimportant background.
However, keeping in mind that  these diagrams are operative
for all three neutrino flavours, it is worthwhile to suppress
this contribution, especially for lower energies. This is best done
by realizing that diagrams 1 \& 2 of Fig.~\ref{fig:feyn_nng}
tend to populate phase space volume corresponding to a (nearly)
real $Z$ production and, hence, can be effectively eliminated by
demanding that
\be
   \left| E_\gamma - \frac{s - m_Z^2}{2 \sqrt{s} }\right|
                 > 5 \Gamma_Z\ ,
                \label{cut:e_gam_win}
\ee
where $\sqrt{s}$ is the centre of mass energy. This obviously means
that the missing mass ($\msl$) is away from $m_Z$ by more than
$5 \Gamma_Z$. With this additional
cut, the total cross section (summed over neutrino flavours) is
\be
\sigma_{\rm SM}(\bar \nu \nu \gamma)
        = \left\{
                \barr{ll}
                0.658 \pb & \qquad \sqrt{s} = 350 \gev \\
                0.848 \pb & \qquad \sqrt{s} = 500 \gev \\
                1.079 \pb & \qquad \sqrt{s} = 800 \gev \ .
                \earr
           \right.
                     \label{nngam_cs}
\ee
Switching on the anomalous couplings, we see immediately that the
cross sections are rather sensitive to these.
In Fig.~\ref{fig:cs_gam}{\em b} we show the cross section as a
function of one anomalous coupling with the other equal to zero.
The sensitivity increases with the center of mass energy (the effect,
expectedly, being more pronounced for $\lg$) since deviations from the
gauge theory prediction leads to a `bad' high energy behaviour.

It is instructive to compare the figures in eq.(\ref{nngam_cs}) with
those for $\sigma_{\rm SM}(W^+ W^-)$, which decreases strongly
with $\sqrt{s}$. In fact, the two are quite comparable for
$\sqrt{s} \sim 800 \gev$. Consequently, the reduction in sensitivity
on account of differences in cross section, becomes less and less
severe as the center of mass energy increases. There is yet another
feature that one should keep in mind. Unlike in $W^+W^-$ production,
the existence of an anomalous coupling affects primarily the
high-energy end of the photon-spectrum. This very concentration
allows one to tune the selection criteria and thereby maximize the
sensitivity. This feature will also hold for $\bar{\nu}\nu Z$
case (Sec.\ref{sec:nnz}).

\subsection{Deriving the constraints}
        \label{sec:constr_gam}
The significant dependence of the total cross section (see
Fig.~\ref{fig:cs_gam}{\em b}) on the TGV would tempt one to
derive bounds from this alone, and this was the approach
taken in refs.\cite{godfrey,abraham}. However, using just the
total cross section is tantamount to discarding a considerable
amount of crucial information,
as is illustrated dramatically in ref.\cite{LEP2}.
Significant improvements can be expected if the phase space
information can be used maximally. While many sophisticated
algorithms like the maximum likelihood method or use of the
optimal variables exist, we desist from using these. Rather,
we choose a simple $\chi^2$ test. Dividing the phase space
into an adequate number of bins, we define
\be
\chi^2 = \sum_{i=1}^{\rm bins}
    \left| \frac{N_{SM}(i) - N_{anom}(i)} {\Delta N_{SM}(i)}
    \right|^2
            \label{chisq}\ ,
\ee
where $N_{SM}$ and $N_{anom}$ are, respectively, the number of
events predicted within the SM  and a theory with an anomalous TGV.
To calculate the number of events, we choose, for the integrated
luminosity,  two typical design values for each choice of center
of mass energy
\be
{\cal L} = \left\{
                \barr{lll}
                10, \: 25 & \fb^{-1} & \qquad \sqrt{s} = 350 \gev \\
                20, \: 50 & \fb^{-1} & \qquad \sqrt{s} = 500 \gev \\
                50, \:120 & \fb^{-1} & \qquad \sqrt{s} = 800 \gev \ .
                \earr
           \right.
                     \label{lumin}
\ee
The error in eq.~(\ref{chisq})
is a combination of statistical and systematic errors
\be
\Delta N = \sqrt{ (\sqrt{N})^2 + (\delta_{syst}N)^2 }
         \label{error}~.
\ee
Since the neutrinos are invisible, the only phase space information
available is the energy and emergence angle of the photon :
($E_\gamma, \cos \theta_\gamma$). Working with this set, we divide
the phase space consistent with the cuts of
eqs.(\ref{cut:th_gam}--\ref{cut:e_gam_win}) into $40 \times 40$ equal
sized bins.
It might be argued that we have resorted to
                an extremely fine binning, especially since
                cross sections are not very large.
                However, the bins are not populated evenly. A large
                fraction of the events are concentrated in relatively
                few bins, leaving the other almost empty.
                To prevent spurious contributions from the latter,
                we discount any bin with less than one event.
                We have also checked that the results do not have
                a very strong dependence on bin cardinality.

Since both the emergence angle and the energy of the photon
should be measurable to a very great
accuracy (particularly within the restricted phase space that we
consider), the main systematic errors originate
from the luminosity measurement and from detector efficiencies.
Since these should not exceed 1\%,
we conservatively assume an overall systematic uncertainty of 2\%,
\ie\  $\delta_{syst} = 0.02$ in eq.(\ref{error}).
The error, for the most part, is thus dominated by the statistical
fluctuations.
\begin{figure}[h]
       \vskip 3.8in\relax\noindent\hskip -1.85in
              \relax{\includegraphics{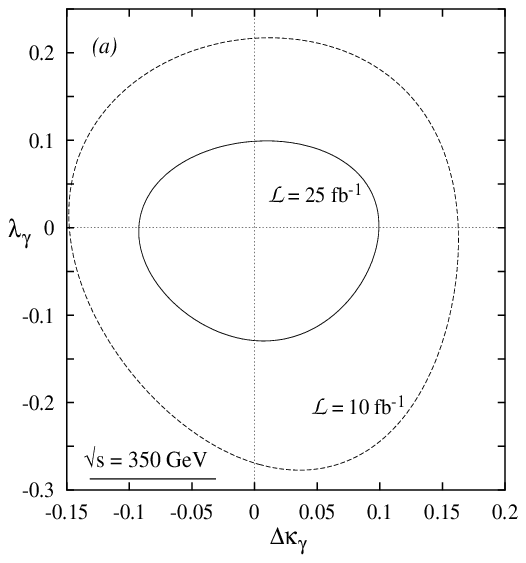}}
       \relax\noindent\hskip 2.125in
              \relax{\includegraphics{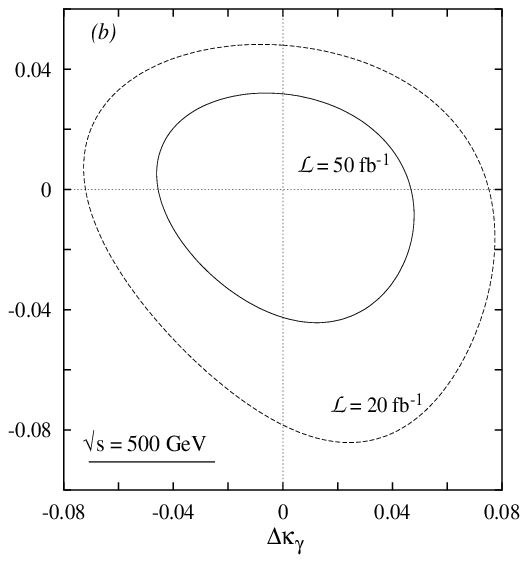}}
       \relax\noindent\hskip 2.125in
              \relax{\includegraphics{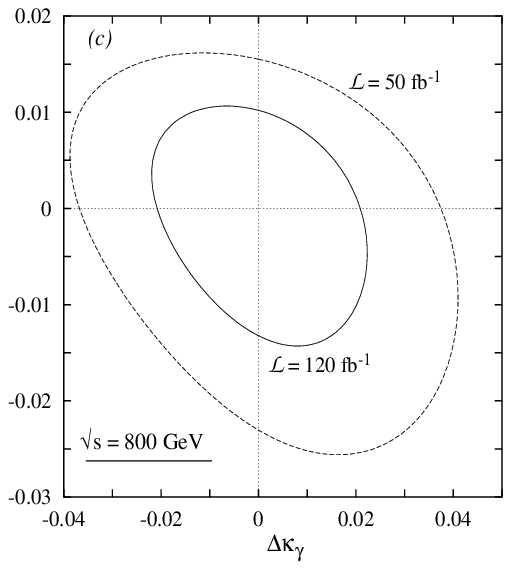}}
        \vspace{-22ex}
 \caption{{\em $\chi^2 = 6$ contours in the $\dkg$--$\lg$ plane for
      various collider configurations. The cuts of
      (\protect\ref{cut:th_gam}--\protect\ref{cut:e_gam_win})
      have been imposed and $ 40 \times 40 $ bins in
      $(\cos \theta_\gamma, E_\gamma)$ used.}}
 \label{fig:kg_lg}
\end{figure}
In Fig.~\ref{fig:kg_lg}, we present the $\chi^2 = 6$ contours in
the $\dkg$--$\lg$ plane for the configurations of eq.(\ref{lumin}).
The interpretation of the figure
is simple.
The areas of the parameter space lying outside each contour
can be ruled out with  95\%\ confidence respectively.
If only one of the parameters is of interest,
irrespective of the values taken by the other,
the half-planes beyond the edges of the contours
can be ruled out at  98.6\%\ C.L. If one of the parameters is known
to take its SM value, the latter confidence levels are valid
for the half-lines bounded by the intersections of the contours with
the axis. Several comments are in order:
\begin{itemize}
  \item Assuming the other coupling to be identically zero, the
        prospective 95\% C.L. bounds are
          \be
             \barr{lll}
               -0.074 < \dkg < 0.079, &
               -0.100 < \lg < 0.091, &
                   \quad {\rm for}\ (350 \gev, 20 \fb^{-1})\\
               -0.036 < \dkg < 0.037, &
               -0.033 < \lg < 0.026, &
                   \quad {\rm for}\ (500 \gev, 50 \fb^{-1})\\
               -0.017 < \dkg < 0.017, &
               -0.010 < \lg < 0.0083, &
                   \quad {\rm for}\ (800 \gev, 120 \fb^{-1})
              \earr
           \ee
        These bounds are one order of magnitude better than those
        obtained in ref.~\cite{abraham} from a consideration of total
        cross section alone.
  \item The bounds on $\lg$ improve much faster with the available
        center of mass energy than those on $\dkg$.
  \item Although not as strong as the bounds expected from $W^+ W^-$
        production, the present ones compare well. In fact, those obtainable
        at the larger $\sqrt{s}$ would probe radiative corrections within
        the MSSM~\cite{MSSM_rad}.
  \item The contours tilt only slightly with respect to the axes,
        showing that there is very little correlation between the two
        parameters $\dkg$ and $\lg$.
  \item Harnessing possible beam polarization is an intriguing thought.
        Two things need to be remembered though. Since the kinematical cuts
        have already eliminated most of the  contribution from the
        $Z$--mediated diagrams (including the interference terms), beam
        polarization affects all relevant diagrams equally. Thus only
        a quantitative change in the contours can be expected. This is in
        marked contrast to the case in the $e^- e^-$ case~\cite{ee}, where
        experiments with different beam polarization settings complement
        each other. Secondly, only left handed polarization would be
        useful\footnote{Thus, if the machine runs only part of the time
             with a left-polarized beam, the bounds will suffer.}.
        Explicit computation with a 90\% left polarized beam and the same
        luminosity shows about 15\% improvement in the bounds.
   \item We have not incorporated initial state radiation effects, which may
        affect performance somewhat.
   \end{itemize}

\section{\boldmath $e^+ e^- \ra \bar \nu \nu Z$}
     \label{sec:nnz}
At the first sight, this is very similar to the case considered in
the previous section, but for the fact that one needs to consider
additional diagrams with the $Z$ radiated from a neutrino. A
similar decomposition to gauge invariant subsets may be done.
Although the expressions for the helicity amplitudes are a little more
complicated owing to the considerable mass of the $Z$, these exist
in the literature~\cite{ambro-mele}. Apparently, all that remains to
be done is to convolute the said expressions with the corresponding decay
($Z \ra f \bar{f}$) matrix amplitudes so as to obtain the relevant
final states. However, such an approximation is valid only in the limit
of exact mass reconstruction.
Since this is unlikely to be as accurate at the LC as at LEP1, one
needs to consider non-resonant production as well. This leads us,
inexorably, to the full $e^+ e^- \ra \nu \bar{\nu} f \bar{f}$ matrix
elements for various choices of $f$.

\begin{figure}[h]
        \vskip 4.4in\relax\noindent\hskip -1.85in
        \relax{\includegraphics{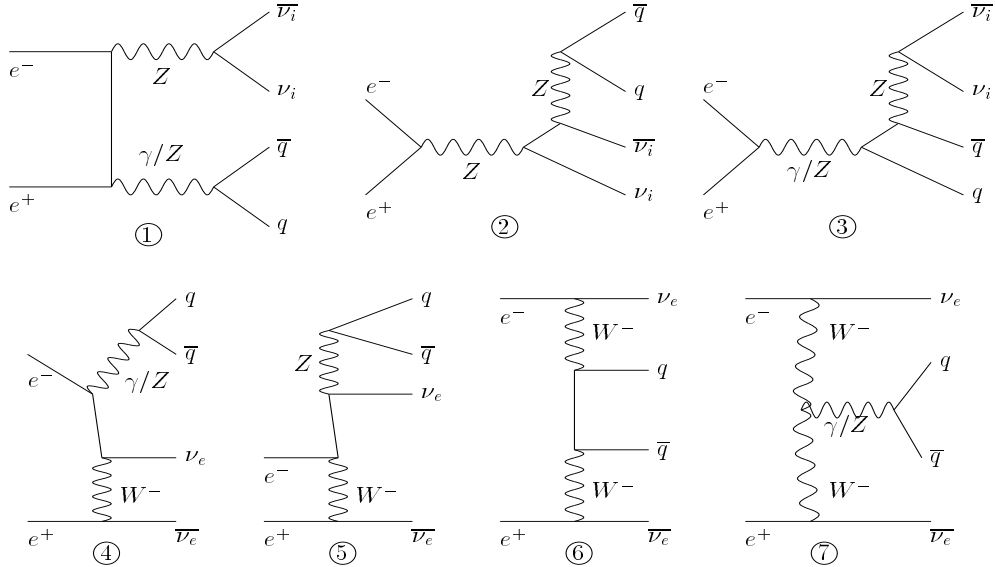}}
        \vspace{-20ex}
 \caption{{\em The different topology classes for Feynman diagrams
               responsible for $e^+ e^- \ra \bar \nu \nu \bar q q$.}}
 \label{fig:feyn_nnqq}
\end{figure}
The Feynman diagrams can be divided into different topology classes
and in  Fig.~\ref{fig:feyn_nnqq},
we exhibit the seven classes for $f = q$.
Of the 19 diagrams, four  each are of types 1,3 \& 4,
two each of types 2,5 \& 7 and finally one of type 6. The diagrams of
type 3 and type 6 would be missed if one had calculated only
$e^+ e^- \ra \bar{\nu} \nu Z$. Again, types 1--3 constitute a
gauge-invariant subset and contribute to all
three neutrino flavours, while types 4--7 are operative for the
electron neutrino only. The diagram of interest is then of type 7.
Note that all six (including $g_1^\gamma$) of the couplings in
eq.(\ref{lagrangian}) can contribute. Does this, then, give a lie to
our claims of the separability of $WW\gamma$ and $WWZ$ couplings?
We shall show below that this is not the case. Indeed, the kinematical
cuts that we will adopt will render the cross section to be very
insensitive to the $WW\gamma$ vertex. Consequently,
any appreciable deviation from that
source could only occur for very large values of
$\dkg$ (or $\lg$) and thus can be ruled out
by the experiment described in the
previous section. This of course does not hold for $\Delta g^1_\gamma$.
However, with the aforementioned cuts, the experiment is sensitive only
to $\Delta g^1_\gamma \sim {\cal O} (10)$,  and thus this particular
form factor can safely be assumed to be vanishing in the course of our
analysis.

To calculate the diagrams we use an appropriately modified
form of the helicity amplitude package MadGraph~\cite{madgraph}.
Extensive counterchecks were performed both through explicit
calculations as well as with CompHEP~\cite{chep}.

\subsection{The signal and the kinematical cuts}
         \label{sec:nnz_cuts}
We choose, in this analysis, to restrict ourselves to $f = q$. Apart from
the fact that the associated cross section is much larger than that for
$f = e,\: \mu, \: \tau$, this has other advantages too. Each of the
last three final states also receive contribution from additional
Feynman diagrams, the most important being $W$ pair-production. As the
latter involves both the $WW\gamma$ and the $WWZ$ vertex (and that too
for a different $q^2$ value), it defeats the avowed objective of our
analysis\footnote{The $\mu^+ \mu^- \nu \bar{\nu}$ final state
                 has, however,
                 been analysed in ref.\protect\cite{godfrey49}.}.

The signal is thus 2 jets accompanied by large missing momentum. For the
jets to be detectable, we require that they have sufficient energy
and be sufficiently away from the beam pipe, \viz,
\be
      E_j > 20 \gev, \quad   20^\circ < \theta_j  < 160^\circ \ .
           \label{cut:e_j,th_j}
\ee
Further, to ensure that
the event selection criteria do not merge the two jets into
a single jet, we demand that the jet-jet angular separation be
large enough~\cite{settles} :
\be
        \theta_{jj} > 20^\circ \ .
             \label{cut:th_jj}
\ee
To enrich the the $\bar \nu \nu Z$ component of the event sample, we
require that the jet-jet invariant mass be sufficiently close to $m_Z$ :
\be
    | m_{jj} - m_Z | < 3 \Gamma_Z \ .
         \label{cut:m_jj}
\ee
This removes, to a very large extent, the contribution from diagram (6),
and from those of diagrams (1,2,4,7) where the exchanged boson is the
photon. We need to ensure that the missing momentum arises on account
of a $\bar \nu \nu$ pair. Therefore, we demand
that the missing momentum ($\pslash$)
have a large enough transverse
component and that its rapidity be small enough :
\be
      \ptsl > 40 \gev, \qquad \eta(\pslash) < 1 \ .
         \label{cut:missmom}
\ee
This, for example, eliminates processes
such as $e^+ e^- \ra \bar q q \gamma$ with the photon vanishing into the
beam pipe. Finally, to eliminate the contributions from diagrams of the
types (1--3), we require that the missing mass
($\msl \equiv \sqrt{\pslash^2} $) be sufficiently away from $m_Z$ :
\be
       |\msl - m_Z | > 5 \Gamma_Z \ .
           \label{cut:missmass}
\ee
\begin{figure}[h]
        \vskip 5in\relax\noindent\hskip -0.7in
        \relax{\includegraphics{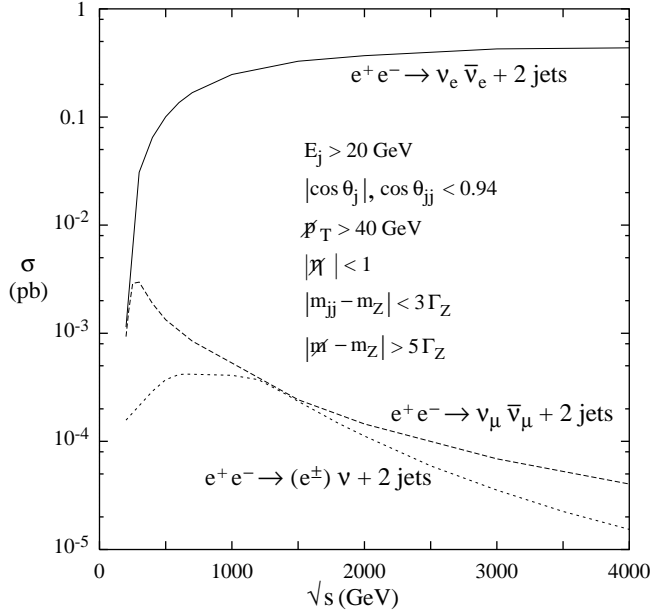}}
        \vspace{-20ex}
 \caption{{\em The variation of the SM cross section with the
               center of mass energy.}}
 \label{fig:cs_Z}
\end{figure}
In Fig.\ref{fig:cs_Z}, we display the variation of
SM  cross section for the $\bar \nu_e \nu_e \bar q q$ process
with the center of mass
energy after the cuts (\ref{cut:e_j,th_j}--\ref{cut:missmass}) have
been imposed. As in the case of $\bar \nu_e \nu_e \gamma$ cross section,
it shows a slow increase with $\sqrt{s}$, a characteristic of the
$t$-channel diagram with the TGV.
While our cross sections are understandably different from those of
ref.\cite{ambro-mele}, we have checked that we reproduce their result
in the proper limit. Also shown in the Fig.\ref{fig:cs_Z} is the
$\bar \nu_\mu \nu_\mu \bar q q$ cross section. As expected, this is
much smaller (the cut eq.(\ref{cut:missmass}) plays a crucial role) and
decreases with the center of mass energy. The total SM cross-section
for the $\bar \nu \nu \bar q q$ (\ie, summed over neutrino and quark
flavours) is thus
\be
\sigma_{\rm SM}(\bar \nu_i \nu_i \bar q q)
        = \left\{
                \barr{ll}
                51.88 \fb & \qquad \sqrt{s} = 350 \gev \\
                103.6 \fb & \qquad \sqrt{s} = 500 \gev \\
                198.4 \fb & \qquad \sqrt{s} = 800 \gev \ .
                \earr
           \right.
                     \label{nnqq_cs}
\ee

We have, until now, not considered other (potentially irreducible)
SM backgrounds to our signal. The most important of these is the
process
\be
  e^+ e^- \ra
       l^\mp {\:\raisebox{-0.2ex}{$\stackrel{\footnotesize (-)}
                                            {\textstyle \nu_l}$}\:}
        \bar q q'
                \label{bkgd_lnqq}
\ee
with the charged lepton moving down the beam pipe.
Understandably, $l = e$
is the main culprit. We determine the corresponding cross section within
the equivalent photon approximation (with any $e^\mp$ within an angle
of  $ 10^\circ $ of the beam considered to be invisible).
The cuts in eqs.(\ref{cut:m_jj} \& \ref{cut:missmom}) play an important role
in reducing this cross section, and were partially designed keeping
this in mind. The resultant cross section is displayed in
Fig.\ref{fig:cs_Z} and is seen to be quite innoccuos.

\begin{figure}[h]
       \vskip 3.8in\relax\noindent\hskip -1.85in
              \relax{\includegraphics{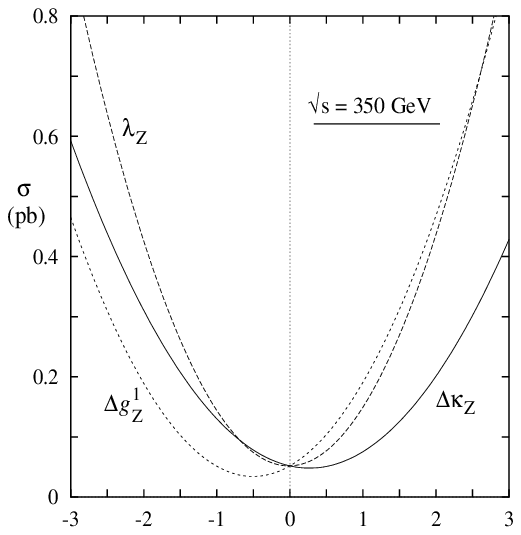}}
       \relax\noindent\hskip 2.125in
              \relax{\includegraphics{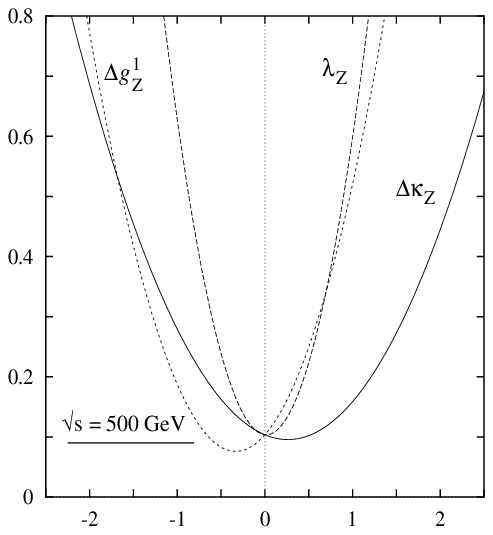}}
       \relax\noindent\hskip 2.125in
              \relax{\includegraphics{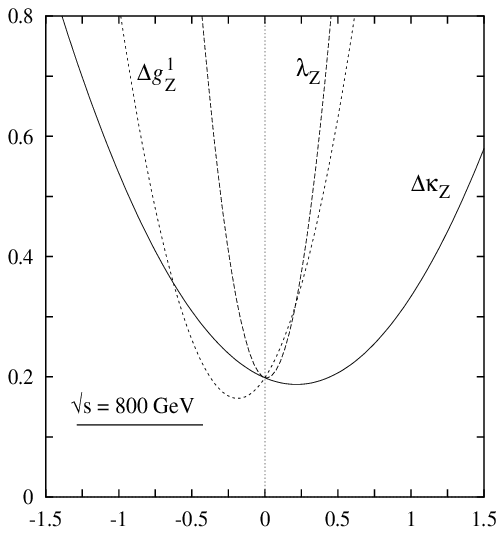}}
        \vspace{-22ex}
 \caption{{\em The variation of the total cross section with individual
               anomalous couplings (the others being zero). The cuts
               (\protect\ref{cut:e_j,th_j}--\protect\ref{cut:missmass})
               have been imposed.
               }}
 \label{fig:xs_Zvar}
\end{figure}
As soon as the anomalous couplings are switched on, the cross sections
change appreciably (see Fig.\ref{fig:xs_Zvar}). As expected, the
deviation is larger for higher center of mass energies, with the
steepest change occurring for $\lZ$. While Ambrosanio \& Mele
(ref.\cite{ambro-mele}) derived their bounds essentially
from plots corresponding to those in Fig.\ref{fig:xs_Zvar}, we shall
attempt to extract more information from the phase space distribution
of the events. To this end, we can follow the procedure of
section \ref{sec:constr_gam}. The situation here is more complicated
though. For one, we have three anomalous couplings instead of two
and the constant $\chi^2$ surfaces are thus two-dimensional.
Secondly, unlike the case in the previos section, we now have quite a few
independent phase space variables. Consequently, we could, in principle,
attempt a more detailed fitting.

We shall, nonetheless, set ourselves a much simpler goal. Instead of
attempting a multivariable fit (which would take us close to the
maximum likelihood method), we shall restrict ourselves to simpler
2-dimensional distributions (\ie, distributions in two independent
kinematical variables), with an effort to identify best such pair for
each detector configuration. We again use
eqs.(\ref{chisq} \& \ref{error}) to calculate the $\chi^2$.
Furthermore, rather than presenting the
constant--$\chi^2$ ellipsoids, we shall only present their projections
on each co-ordinate plane. The interpretation is then akin to that of
section \ref{sec:constr_gam} with the constraint that the third
coupling is identically vanishing.

\subsection{The constraints}
       \label{sec:constr_Z}
As the cross sections are significantly smaller than those in
section \ref{sec:nngamma}, we adopt coarser binnings, especially for
the lower energy options. While distributions in different kinematical
variables are sensitive to the three couplings to different extent, we
have restricted ourselves to one particular pairing for each $\sqrt{s}$,
our choice being
\be
\barr{lll}
\sqrt{s} = 350 \gev \ : & \left( min(|\cos \theta_1|,|\cos \theta_2|),
                             E_1 + E_2
                       \right)
                     & 20 \times 20 \; {\rm bins}  \ ,
          \\[1ex]
\sqrt{s} = 500 \gev \ : & \left( min(|\cos \theta_1|,|\cos \theta_2|),
                             \: {\rm corresponding} \: E_i
                       \right)
                     & 30 \times 30 \; {\rm bins} \ ,
          \\[1ex]
\sqrt{s} = 800 \gev \ : & \left( max(|\cos \theta_1|,|\cos \theta_2|),
                             \: {\rm corresponding} \: E_i
                       \right)
                     & 40 \times 40 \; {\rm bins} \ ,
\earr
        \label{Z_binning}
\ee
where the subscripts refer to the jets.
In Figs.\ref{fig:kz_lz}({\em a--i}) we display the
95\%~C.L. exclusion contours for the three different pairings of the
anomalous couplings. The interpretation is the same as in
section \ref{sec:constr_gam}.

The main conclusions can be summarised as follows :
\begin{itemize}
  \vspace*{-2.5ex}
 \item Assuming the other two couplings to be identically zero, the
        95\% C.L. bounds, for typical
$(\sqrt{s}/{\rm GeV}, {\cal L}/{\rm fb}^{-1} )$ combinations, would be
     \be
     \barr{rccc}
      (350,25): \hspace*{1em} &
           -75 < \bdkZ < 90, &
           -104  < \blZ  < 79, &
           -40 < \bdgZ < 37, \\[1.2ex]
      (500,50): \hspace*{1em} &
           -35  < \bdkZ < 38, &
           -5.6  < \blZ  < 5.5, &
           -15  < \bdgZ < 14, \\[1.2ex]
      (800,120): \hspace*{1em} &
           -10 < \bdkZ < 9.9, &
           -2.1 < \blZ  < 2.1, &
           -4.3 < \bdgZ < 4.2,
     \earr
     \ee
where $\bdkZ \equiv \dkZ / 1000$ etc.
\begin{figure}[h]
       \vskip 3.45in\relax\noindent\hskip -1.7in
              \relax{\includegraphics{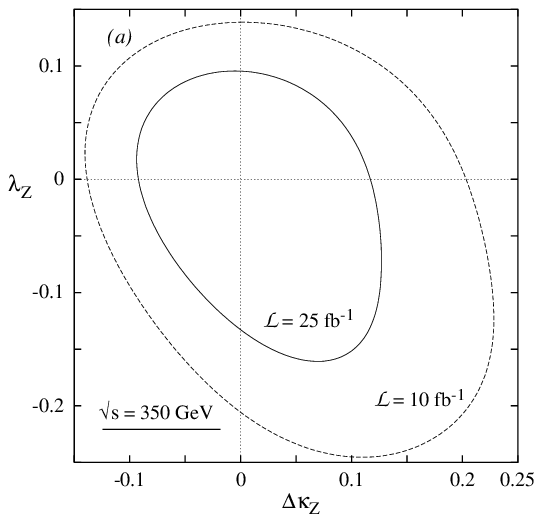}}
       \relax\noindent\hskip 2.1in
              \relax{\includegraphics{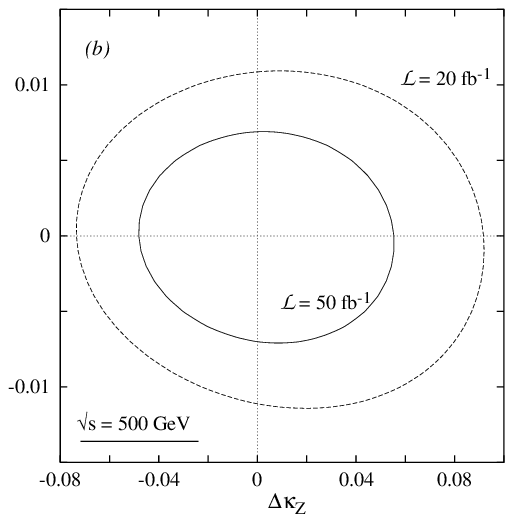}}
       \relax\noindent\hskip 2.1in
              \relax{\includegraphics{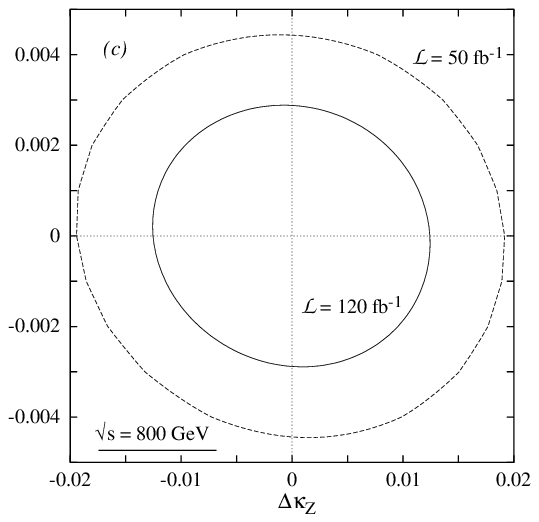}}
       \vskip 1.95in\relax\noindent\hskip -1.7in
              \relax{\includegraphics{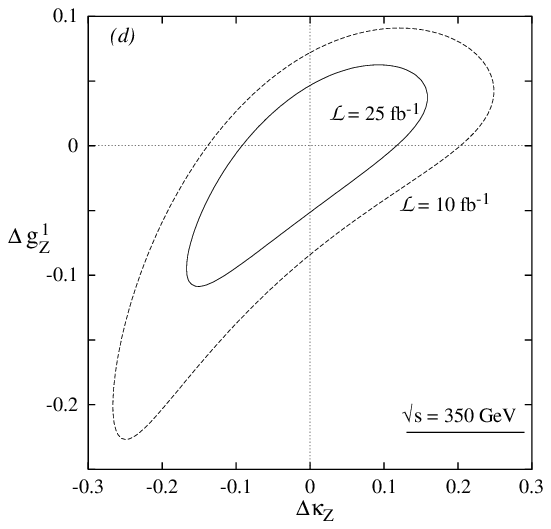}}
       \relax\noindent\hskip 2.1in
              \relax{\includegraphics{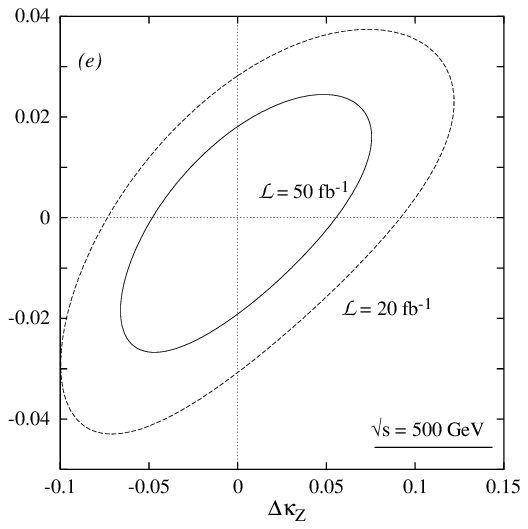}}
       \relax\noindent\hskip 2.1in
              \relax{\includegraphics{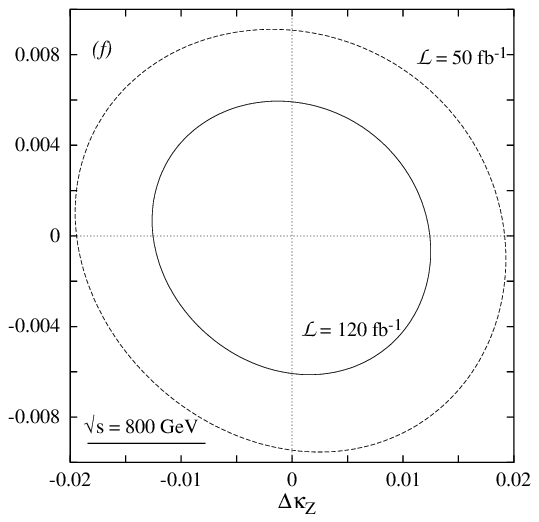}}
       \vskip 1.95in\relax\noindent\hskip -1.7in
              \relax{\includegraphics{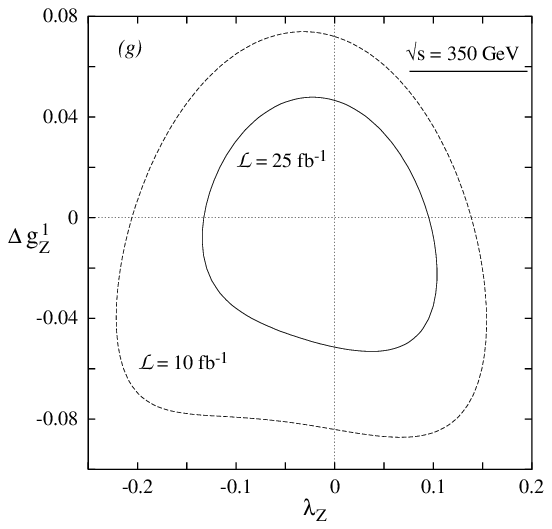}}
       \relax\noindent\hskip 2.1in
              \relax{\includegraphics{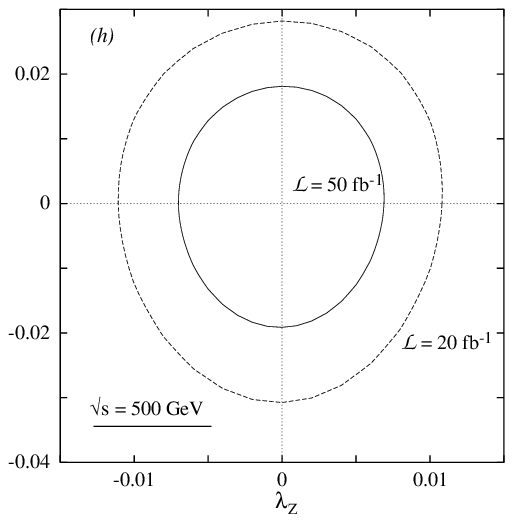}}
       \relax\noindent\hskip 2.1in
              \relax{\includegraphics{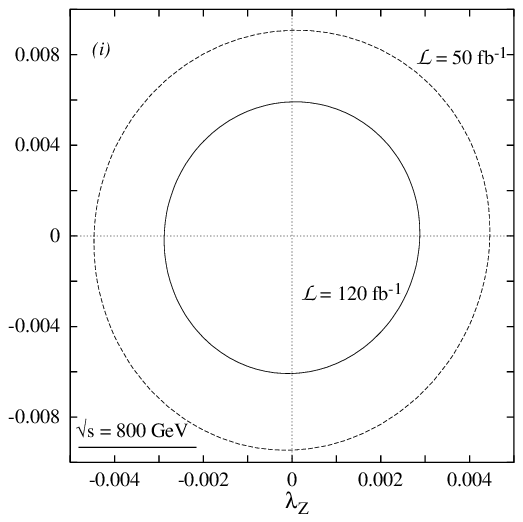}}
        \vspace{-22ex}
 \caption{{\em Projections of the $\chi^2 = 6$ contours on the three
      coordinate planes for
      various collider configurations. The cuts
           (\protect\ref{cut:e_j,th_j}--\protect\ref{cut:missmass})
               have been imposed and the binnings of eq.(\ref{Z_binning})
               adopted.
               {\em (a--c)} on $\dkZ$--$\lZ$ plane;
               {\em (d--f)} on $\dkZ$--$\dgZ$ plane;
               {\em (g--i)} on $\lZ$--$\dgZ$ plane.
          }}
 \label{fig:kz_lz}
\end{figure}

 \item The bounds are thus quite comparable to those expected from
       $W$ pair production at the same machine. Exploration of radiative
       contributions to this vertex thus becomes possible. \\
       The second generation machines (higher center of mass energy and
       luminosity) will lead to a significant improvement in the
       constraints, with the largest improvement occuring for the
       $\lZ$.

 \item The most discernible correlations are for the ($\dkZ$, $\dgZ$)
       pair. This is to be expected as $ \dgZ = \dkZ$ implies that
       only the overall $WWZ$ coupling has been changed. \\
       This correlation disappears for higher $\sqrt{s}$ as the
       non-gauge nature of the theory becomes more apparent.

 \item Inclusion of the initial state radiation effects (neglected
       in this analysis) is expected to degrade the results somewhat.
       On the other hand, the use of the full phase space information
       (whether maximum likelihood or optimal variables) can be
       expected to improve the bounds.

  \item As in section \ref{sec:constr_gam}, beam polarization is not
       a priority.

\end{itemize}

\section{Summary}
   We have shown that the processes
$e^+e^- \rightarrow \bar{\nu}\nu\gamma$
and $\bar{\nu}\nu Z(\rightarrow \bar{q}q)$ can be very useful in
disentangling  the anomalous $WW\gamma$ and $WWZ$ couplings. We have
included all tree level processes that lead to these final states.
The signal to background ratio is improved by imposing a set of
kinematical cuts. Contrary to naive expectations, the relevant
cross sections are not severely suppressed in comparison with those
for $W$-pair-production. In addition, the effect of possible
anomalous couplings is concentrated largely in the high-energy
end of the $\gamma$ ($\bar{q} q$) spectrum thus allowing for a
more efficient extraction of information.

Exploiting the
full phase space knowledge of the cross sections, one can derive bounds
that are comparable to those expected from  $W^+W^-$ production.
Apart from disentangling the $WW\gamma$ and the $WWZ$ couplings,
this measurement has the added advantage of being performed
at a different momentum transfer. Coupled with the $W^+ W^-$
measurements, these then
offer the possibility of studying
the form-factor nature of these couplings.

\noindent
{\bf Acknowledgements :}
We would like to thank R.~Settles, G.~Wilson and P.M.~Zerwas
for many discussions and continous encouragement. JK thanks M. Krawczyk
for useful discussion and help.

\end{document}